\documentclass{emulateapj}

\newcommand{\etal}    {{\it et al.~}}

\shorttitle{IR Morphology of SNRs}
\slugcomment{Accepted by ApJL}
\shortauthors{Peters, C., \etal}

\newcommand{\ltsima}{$\; \buildrel < \over \sim \;$}
\newcommand{\simlt}{\lower.5ex\hbox{\ltsima}}

\newcommand{\ls}{{_<\atop^{\sim}}}

\newcommand{\gs}{{_>\atop^{\sim}}}

\newcommand{\ptwo}{$P_{2}/P_{0}$}
\newcommand{\pthree}{$P_{3}/P_{0}$}

\begin{document}

\title{Constraining explosion type of young supernova remnants using 24 $\mu$m emission morphology}

\author{
Charee L. Peters\altaffilmark{1,5}, Laura A.Lopez\altaffilmark{2,6,7}, Enrico Ramirez-Ruiz\altaffilmark{3}, Keivan G. Stassun\altaffilmark{4,1}, Enectali Figueroa-Feliciano\altaffilmark{2} 
}

\email{charee.l.peters@vanderbilt.edu} 

\altaffiltext{1}{Department of Physics, Fisk University, 1000 17th Ave N Nashville, TN 37208, USA; charee.l.peters@vanderbilt.edu.} 
\altaffiltext{2}{MIT-Kavli Institute for Astrophysics and Space Research, 77 Massachusetts Avenue, 37-664H, Cambridge MA 02139, USA}
\altaffiltext{3}{Department of Astronomy and Astrophysics, University of California Santa Cruz, 1156 High Street, Santa Cruz, CA 95060, USA}
\altaffiltext{4}{Department of Physics and Astronomy, Vanderbilt University, 6301 Stevenson Center Ln., Nashville, TN 37235, USA}
\altaffiltext{5}{Fisk-Vanderbilt Master's-to-PhD Bridge Program Fellow}
\altaffiltext{6}{NASA Einstein Fellow}
\altaffiltext{7}{Pappalardo Fellow in Physics}

\begin{abstract}

Determination of the explosion type of supernova remnants (SNRs) can be challenging, as SNRs are hundreds to thousands of years old and supernovae (SNe) are classified based on spectral properties days after explosion. Previous studies of thermal X-ray emission from Milky Way and Large Magellanic Cloud (LMC) SNRs have shown that Type Ia and core-collapse (CC) SNRs have statistically different symmetries, and thus these sources can be typed based on their X-ray morphologies. In this paper, we extend the same technique, a multipole expansion technique using power ratios, to infrared (IR) images of SNRs to test whether they can be typed using the symmetry of their warm dust emission as well. We analyzed archival {\it Spitzer Space Telescope} Multiband Imaging Photometer (MIPS) 24 $\mu$m observations of the previously used X-ray sample, and we find that the two classes of SNRs separate according to their IR morphologies. The Type Ia SNRs are statistically more circular and mirror symmetric than the CC SNRs, likely due to the different circumstellar environments and explosion geometries of the progenitors. Broadly, our work indicates that the IR emission retains information of the explosive origins of the SNR and offers a new method to type SNRs based on IR morphology. 

\end{abstract}


\section{Introduction} \label{sec:introduction}

Supernovae (SNe) can be divided into two physically distinct classes: Type Ia and core-collapse (CC: Types Ib, Ic, and II) explosions. Type Ia SNe are thought to arise from the thermonuclear detonation of a white dwarf; CC SNe occur after the gravitational collapse of a massive star ($\gs$8 $M_{\sun}$; \citealt{woosley05}). Observationally, SNe are classified using their optical spectra near maximum brightness, days to weeks after the explosions \citep{filippenko97}. Thus, alternative means are necessary to constrain the explosion types of young supernova remnants (SNRs), the thermodynamical structures imprinted in the ISM hundreds to thousands of years after the SNe. In the literature, authors have employed numerous lines of evidence to elucidate the nature of SNR progenitors (see \citealt{vinkreview} for a review): abundance ratios of shock-heated ejecta, a coincident neutron star or pulsar, spectra of SN light echoes, and environment (e.g., a nearby OB association or emission from massive star wind material). Each of these indicators has associated challenges or shortcomings: for example, Type Ia SNRs have been found in star-forming (e.g., N103B: \citealt{lewis03}) or dense environments (e.g., Kepler: \citealt{reynolds07}), and SNRs may have chance alignments with pulsars \citep{kaspi98}. Therefore, complementary means to classify SNRs are useful to confirm the progenitors' nature, and accurate typing of SNRs is critical to probe the recent, local population of SNe and their physical properties. 

Recently, \citet[hereafter L09a and L11, respectively]{lopez09b,lopez11} have demonstrated that the X-ray morphology of SNRs can also be used to constrain explosion types. In particular, L09a and L11 performed a quantitative analysis of {\it Chandra} X-ray Observatory images of young ($\ls$25000 years old) Milky Way and Large Magellanic Cloud (LMC) SNRs, and they showed that the X-ray line and thermal bremsstrahlung emission of Type Ia SNRs are more circular and mirror symmetric than core-collapse SNRs. The observed differences in morphologies are sufficiently large to permit a clean separation of the two explosion classes. These results arise due to the different geometries of the explosion mechanisms and circumstellar environments of Type Ia and CC SNRs. 

\begin{deluxetable*}{lllccccccl}[t]
\tablecolumns{8} \tablewidth{0pc} \tabletypesize{\footnotesize}
\setlength{\tabcolsep}{0.05in} \renewcommand{\arraystretch}{1.0}
\tablewidth{0pt}
\tablecaption{Sources, Sorted by Age}
\footnotesize
\tablehead{\colhead{\#} & \colhead{Source} & \colhead{AORKEY} & \colhead{Age\tablenotemark{a}} & \colhead{Distance} & \colhead{Radius\tablenotemark{b}} & \colhead{Evidence of} & \colhead{\ptwo} & \colhead{\pthree} & \colhead{References} \\ 
\colhead{} & \colhead{} & \colhead{} & \colhead{(years)} & \colhead{(kpc)} & \colhead{(pc)} & \colhead{Explosion Type\tablenotemark{c}} & \colhead{($\times 10^{-6}$)} & \colhead{($\times 10^{-7}$)} & \colhead{}}
\startdata
\cutinhead{Type Ia Sources} 
1 & 0509$-$67.5 & 11542016 & 350--450 & 50 & 9.42 & S, L & $1.78^{+1.77}_{-1.60}$ & $4.08^{+2.52}_{-3.18}$ & 1, 2 \\
2 & Kepler & 10914816 & 405 & 5.0 & 4.39 & S & $2.28^{+0.10}_{-0.11}$ & $2.10^{+0.15}_{-0.18}$ & 3 \\
3 & 0519$-$69.0 & 11542272 & 400--800 & 50 & 10.49 & S, L & $2.19^{+1.10}_{-1.13}$ & $0.81^{+0.55}_{-0.65}$ & 4, 5 \\
4 & N103B & 11531520 & 860 & 50 & 10.85 & S & $1.52^{+0.34}_{-0.34}$ & $1.60^{+0.58}_{-0.65}$ & 6 \\
5 & DEM L71 & 11521536 & $\sim$4360 & 50 & 15.17 & S & $6.61^{+3.35}_{-4.30}$ & $4.95^{+3.71}_{-4.05}$ & 7 \\ 
\cutinhead{Core-collapse Sources}
6 & Cas A & 8099840 & 309--347 & 3.4 & 4.27 & N, S, L & $2.74^{+0.04}_{-0.03}$ & $5.47^{+0.06}_{-0.05}$ & 8 \\
7 & W49B & 11021824 & $\sim$1000 & 8.0 & 7.31 & S & $27.3^{+0.4}_{-0.5}$ & $20.1^{+0.5}_{-0.6}$ & 9 \\
8 & G15.9$+$0.2 & 15575808 & $\sim$1000 & 8.5 & 9.24 & N, S & $35.4^{+1.7}_{-1.6}$ & $42.3^{+2.9}_{-2.9}$ & 10 \\
9 & G11.2$-$0.3 & 15629824 & 1623 & 5.0 & 4.50 & N, S & $8.30^{+0.32}_{-0.29}$ & $6.37^{+0.44}_{-0.46}$ & 11 \\
10 & Kes 73 & 15651840 & 500--2200 & 8.0 & 7.82 & N, S & $21.7^{+0.7}_{-0.7}$ & $14.5^{+0.7}_{-0.8}$ & 12 \\
11 & N132D & 11527168 & $\sim$3150 & 50 & 25.23 & S & $0.64^{+0.23}_{-0.25}$ & $63.7^{+4.2}_{-4.2}$ & 13 \\
12 & G292.0$+$1.8 & 23088896 & $\sim$3300 & 6.0 & 9.69 & N, S & $28.5^{+1.3}_{-1.5}$ & $10.4^{+1.4}_{-1.2}$ & 14 \\ 
13 & 0506$-$68.0 & 11533824 & $\sim$4600 & 50 & 16.88 & N, S & $13.3^{+2.4}_{-2.6}$ & $35.6^{+7.0}_{-6.5}$ & 15 \\
14 & N49B & 11526912 & 10000 & 50 & 28.14 & S & $30.1^{+7.7}_{-6.4}$ & $105^{+24}_{-20}$ & 16 \\
15 & B0453$-$685 & 11532288 & 13000 & 50 & 23.20 & N, S & $159^{+81}_{-88}$ & $94.5^{+90.7}_{-74.5}$ & 17 \\
\enddata
\tablenotetext{a}{Ages are uncertain for the sources which are not historical SNRs nor have detected light echoes.}
\tablenotetext{b}{Radius $R$ selected to enclose the entire source in the 24 $\mu$m image; $R$ is determined assuming the distances given in the table.} 
\tablenotetext{c}{Evidence of explosion type: S = spectral properties (e.g. Prominent line emission associated with Type Ia SNe (Fe L emission) or core-collapse SNe (enhanced O, Mg, and/or Ne); L = light echoes; N = detection of a neutron star} 
\tablerefs{[1]~\cite{bad08}; [2]~\cite{rest08}; [3]~\cite{reynolds07}; [4]~\cite{rest05}; [5]~\cite{hughes95}; [6]~\cite{lewis03}; [7]~\cite{hughes03}; [8]~\cite{hwang04}; [9]~\cite{lopez13}; [10]~\cite{reynolds06}; [11]~\cite{kaspi01}; [12]~\cite{gotthelf}; [13]~\cite{bork07}; [14]~\cite{hughes01}; [15]~\cite{hughes06}; [16]~\cite{park03}; [17]~\cite{gaensler03}}
\label{table:sources}
\end{deluxetable*}

In this paper, we extend the symmetry technique of L09a and L11 to {\it Spitzer Space Telescope} infrared (IR) images of young SNRs to test whether Type Ia and CC SNR morphologies are distinct at these wavelengths as well. The IR emission of SNRs arises from several mechanisms: e.g., stochastically and thermally heated dust \citep{reach06,andersen11,pin11}, atomic/molecular lines (e.g., \citealt{arendt99}), polycyclic aromatic hydrocarbons (PAHs; e.g., \citealt{tappe12}), and synchrotron radiation \citep{rho03}. Previous IR surveys of SNRs with {\it IRAS} (e.g., \citealt{saken92}) and with {\it Spitzer Space Telescope} (e.g., \citealt{bork06,williams06,pin11}) have shown that a substantial component of the mid-IR emission comes from dust collisionally heated by the hot plasma generated by SNR shocks \citep{draine81,dwek87}. This conclusion is supported by the morphological similarities between 24 $\mu$m emission and X-ray features in {\it Chandra} images \citep{bork06,williams06,pin11}, where the IR appears to trace the SNR blast waves seen in X-rays. Thus, we may expect a similar morphological difference between Type Ia and CC SNRs at IR wavelengths as in X-ray wavelengths. 

The organization of this paper is as follows. We discuss the sample and data used in our analyses in Section~\ref{sec:data}. We outline the power-ratio method (PRM), a multipole expansion technique, employed for our symmetry analyses in Section~\ref{sec:methods}. In Section~\ref{sec:results&discuss}, we present the results and discuss the implications regarding the IR morphological differences of Type Ia and CC SNRs. Section~\ref{sec:summary} summarizes the findings of this work. 

\section{Sample \& Data Preparation} \label{sec:data}

\begin{figure*}[t]
\begin{center}
\includegraphics[width=0.95\textwidth]{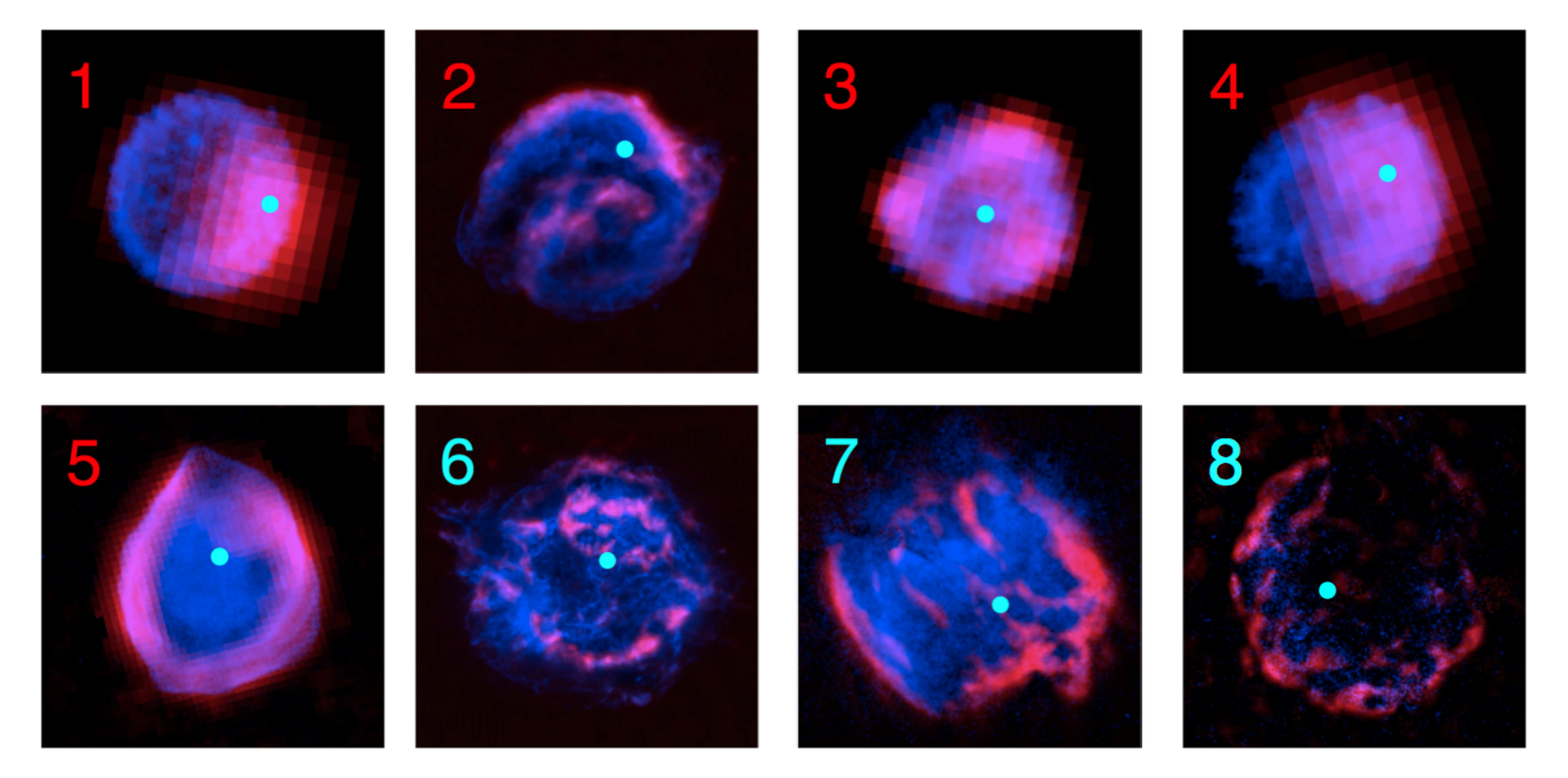}
 \caption{Images of {\it Spitzer} 24 $\mu$m emission (red) and {\it Chandra} soft X-ray (0.5--2.1 keV) emission (blue) of 8 SNRs listed in Table~\ref{table:sources} with background and point sources removed. Numbers correspond to those listed in Column 1 of Table~\ref{table:sources}. Red numbers are Type Ia SNRs; cyan numbers are CC SNRs. The blue dots in each image mark the centroid of the 24 $\mu$m emission used in our power-ratio method analyses.}
\label{fig:24emissionIa}
\end{center}
\end{figure*}

\begin{figure*}[t]
\begin{center}
\includegraphics[width=0.95\textwidth]{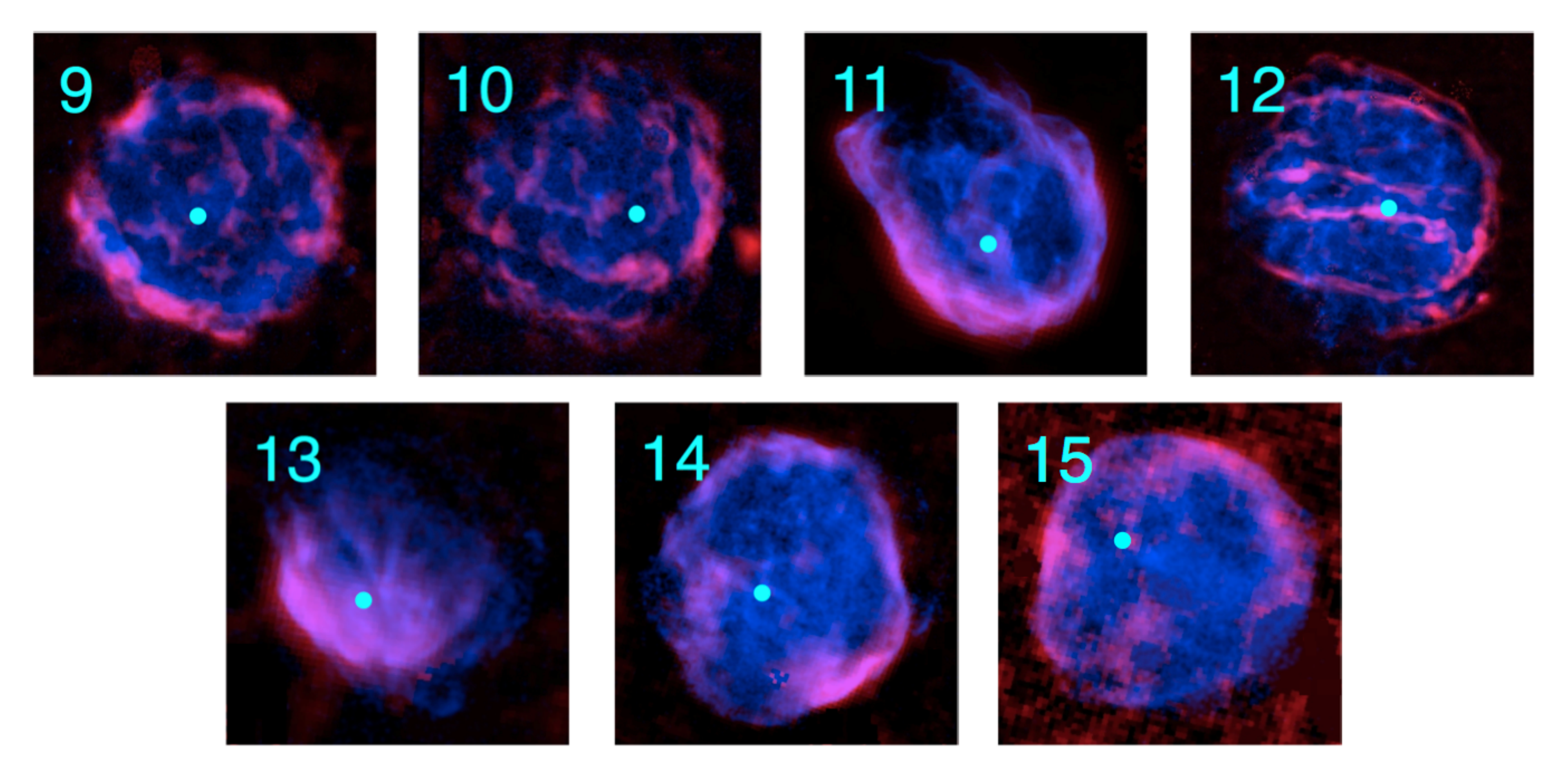}
 \caption{Same as Figure~\ref{fig:24emissionIa} for SNRs numbered 9--15 in Table~\ref{table:sources}.}
\label{fig:24emissionCC}
\end{center}
\end{figure*}

We analyzed archival {\it Spitzer} Multiband Imaging Photometer (MIPS; \citealt{rieke04}) observations of the 15 SNRs listed in Table~\ref{table:sources} and shown in Figures~\ref{fig:24emissionIa} and \ref{fig:24emissionCC}. We have selected SNRs in both the Milky Way and the LMC which have resolved, extended emission in the {\it Spitzer} images. Our IR sample is a subset of the 24 SNRs in the L11 X-ray sample. We have excluded 9 L11 sources which are not detected in the mid- or far-IR, are not well separated from the IR background, or are not available publicly in the {\it Spitzer} archive. The 15 SNRs have been observed with {\it Spitzer} MIPS, either through targeted observations or via surveys of the Galactic plane \citep{churchwell09} and of the LMC \citep{meixner}. In our sample, 5 of the sources have been shown to have likely originated from Type Ia SNe, while 10 are considered to have been from CC SNe. The basis for these SN classifications (and associated references) is listed in Table~\ref{table:sources}. 

We opted to measure the {\it Spitzer} IR morphologies of our sample at 24 $\mu$m for several reasons. At the near-IR wavelengths of the Infrared Array Camera (IRAC: \citealt{fazio04}), the emission of galaxies is dominated by starlight (at $\ls$5 $\mu$m: \citealt{helou04}) and by PAHs (at 8$\mu$m: \citealt{dl07}). At mid-IR wavelengths, several emission mechanisms contribute (see Section \ref{sec:introduction}), but the dominant component of the IR radiation in the ISM is from warm dust at 24 $\mu$m and 70 $\mu$m \citep{dl07}. Based on previous {\it Spitzer} surveys of SNRs at 24 $\mu$m and 70 $\mu$m, the former wavelength has the highest detection rate, with 32\% of the Green SNR Catalog \citep{green09} detected at 24 $\mu$m \citep{pin11}. Figures~\ref{fig:24emissionIa} and \ref{fig:24emissionCC} shows the MIPS 24 $\mu$m images of the 15 SNRs that comprise our sample together with the {\it Chandra} X-ray images (from L11). 

Data were downloaded from the {\it Spitzer} Heritage Archive\footnote{http://archive.spitzer.caltech.edu/} and processed through the {\it Spitzer} MOSaicker and Point source EXtractor (MOPEX) GUI 18.5.0. We employed MOPEX to remove the background from each source by following the Overlap, Mosaic, and APEX 1frame template pipelines. 

\section{Methods} \label{sec:methods}

Following the data preparation, we applied the power-ratio method (PRM) to compare the IR morphologies of Type Ia and CC SNRs. The PRM quantifies the asymmetry in surface brightness distributions via calculation of multipole moments in a circular aperture. The PRM was used to characterize the X-ray morphologies of galaxy clusters \citep{b96,j05}. Subsequently, the technique was extended to {\it Chandra} observations of SNRs to quantify the relative distribution of elements in individual SNRs \citep{lopez09a} and to compare the morphologies of X-ray line and thermal bremsstrahlung emission in Type Ia and CC SNRs (L09a and L11). 

The PRM employs a two-dimensional multipole expansion, which is derived similarly to the 2D expansion of a gravitational potential within an enclosed radius $R$:

\begin{eqnarray}
\lefteqn{\Psi(R,\phi) = -2Ga_0\ln\left({1 \over R}\right)-2G }
\nonumber \\ & & \times \sum^{\infty}_{m=1} {1\over m
R^m}\left(a_m\cos m\phi + b_m\sin
m\phi\right), \label{eqn.multipole}
\end{eqnarray}

\noindent
where $G$ is the gravitational constant and the moments $a_m$ and $b_m$ are

\begin{eqnarray}
a_m(R) & = & \int_{R^{\prime}\le R} \Sigma(\vec x^{\prime})
\left(R^{\prime}\right)^m \cos m\phi^{\prime} d^2x^{\prime}, \nonumber \\
b_m(R) & = & \int_{R^{\prime}\le R} \Sigma(\vec x^{\prime})
\left(R^{\prime}\right)^m \sin m\phi^{\prime} d^2x^{\prime}, \nonumber
\end{eqnarray}

\noindent
Here, the position vector $\vec x^{\prime} = (R^{\prime},\phi^{\prime})$, and $\Sigma$ is the surface mass density, or for our analysis, the 24 $\mu$m surface brightness.

The powers of the multipole expansion are obtained through integration of the magnitude of $\Psi_m$ (the \textit{m}th term in the multipole expansion) over a circle of radius $R$:

\begin{equation}
P_m(R)={1 \over 2\pi}\int^{2\pi}_0\Psi_m(R, \phi)\Psi_m(R, \phi)d\phi.
\end{equation}

\noindent
The above equation reduces to: 

\begin{eqnarray}
P_0 & = & \left[a_0\ln\left(R\right)\right]^2 \nonumber \\
P_m & = & {1\over 2m^2 R^{2m}}\left( a^2_m + b^2_m\right) 
\end{eqnarray}
		
\noindent		
To account for the different fluxes of sources, we divide the powers by $P_{0}$ to form the power ratios, $P_{m}/P_{0}$. 

The morphology and shape of the infrared emission is reflected through the moments $a_{m}$ and $b_{m}$ and the power ratios $P_{m}/P_{0}$. Higher-order terms are probing the asymmetry at successively smaller scales. We have set the origin from which we draw our radius $R$ at the centroid of the 24 $\mu$m surface brightness, so the dipole power, $P_{1}$, approaches zero. Higher-order powers' values quantify the morphology of the emission. For our analysis, we focus on the quadrupole power ratio $P_{2}/P_{0}$ and the octupole power ratio $P_{3}/P_{0}$. The quadrupole power ratio, $P_{2}/P_{0}$, is a measurement of the ellipticity of the distribution. Large $P_{2}/P_{0}$ values are derived for sources with elliptical/elongated morphologies; conversely, circular distributions give small $P_{2}/P_{0}$ values. The octupole power ratio, $P_{3}/P_{0}$, characterizes the mirror asymmetry of a source. Asymmetric or non-uniform surface brightness distributions yield large $P_{3}/P_{0}$, while mirror symmetric sources have small $P_{3}/P_{0}$.

\begin{figure}
\centering
\includegraphics[width=\columnwidth]{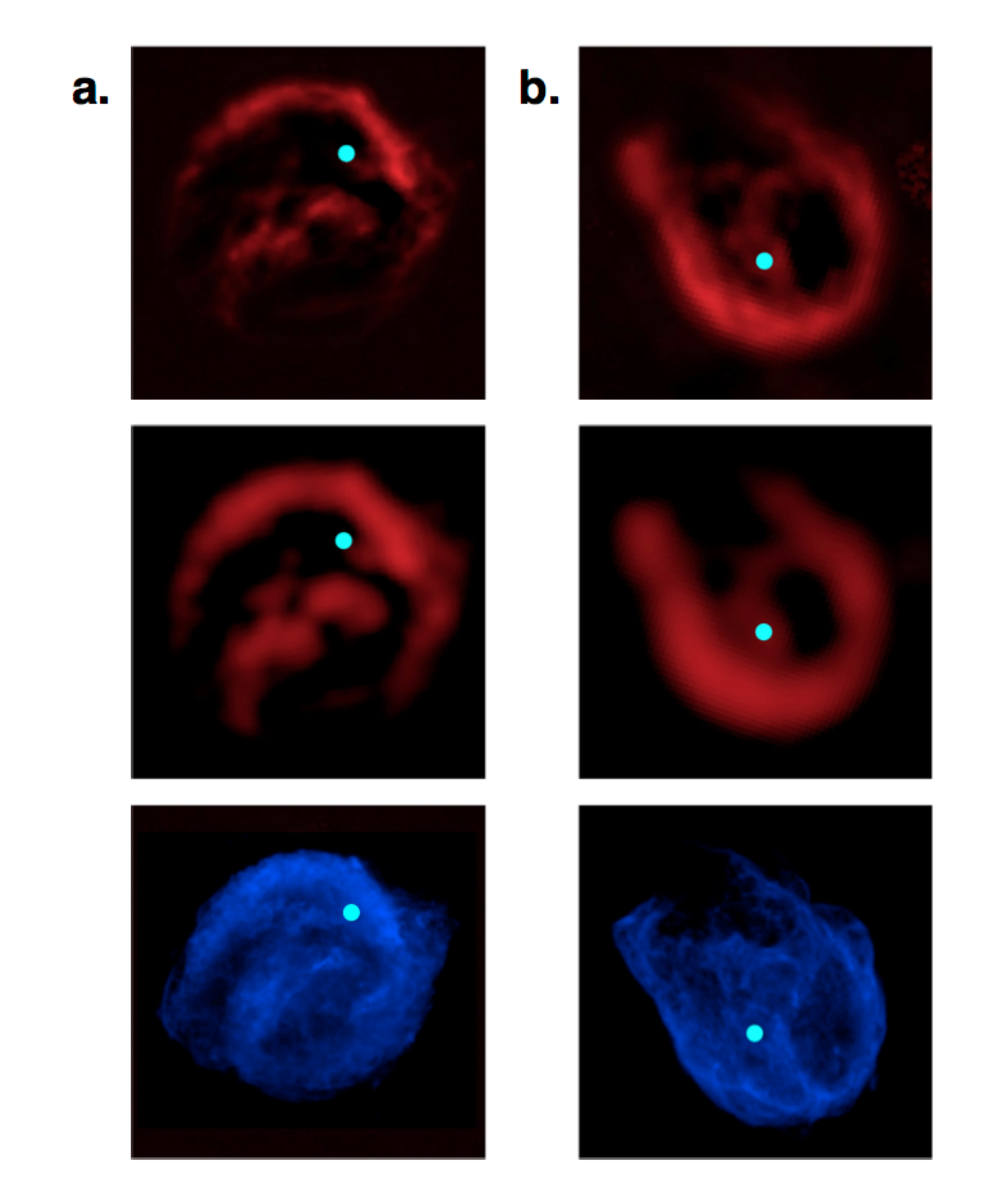}
\caption{{\it a.} 24 $\mu$m image (top) and an example 24 $\mu$m mock image (middle) of the Type Ia SNR Kepler, after background and point source removal. For comparison, the bottom image shows the {\it Chandra} X-ray image of Kepler. {\it b.} The same as {\it a} but for the CC SNR N132D. The blue dots in each image mark the centroid of the 24 $\mu$m emission used in our power-ratio method analyses.}
\label{fig:mockimages}
\end{figure}

To estimate the uncertainty in the derived power ratios, we used a Monte Carlo approach similar to that of \cite{lopez09a,lopez11}. In particular, we took our background-subtracted images and smoothed out noise. Then, we put noise back into the image by assuming each pixel intensity is the mean of a Poisson distribution and selecting a new intensity from that distribution. We repeated this process 100 times for each SNR IR image, thus creating 100 mock images of each source; Figure~\ref{fig:mockimages} gives example mock images of two sources, N132D and Kepler, following this procedure. Then, we calculated the power ratios of the 100 images, and the 1-$\sigma$ confidence limits were determined based on the values of the sixteenth highest and lowest power ratios obtained from the 100 mock images. 

\section{Results and Discussion} \label{sec:results&discuss}

\begin{figure}
\centering
\includegraphics[width=\columnwidth]{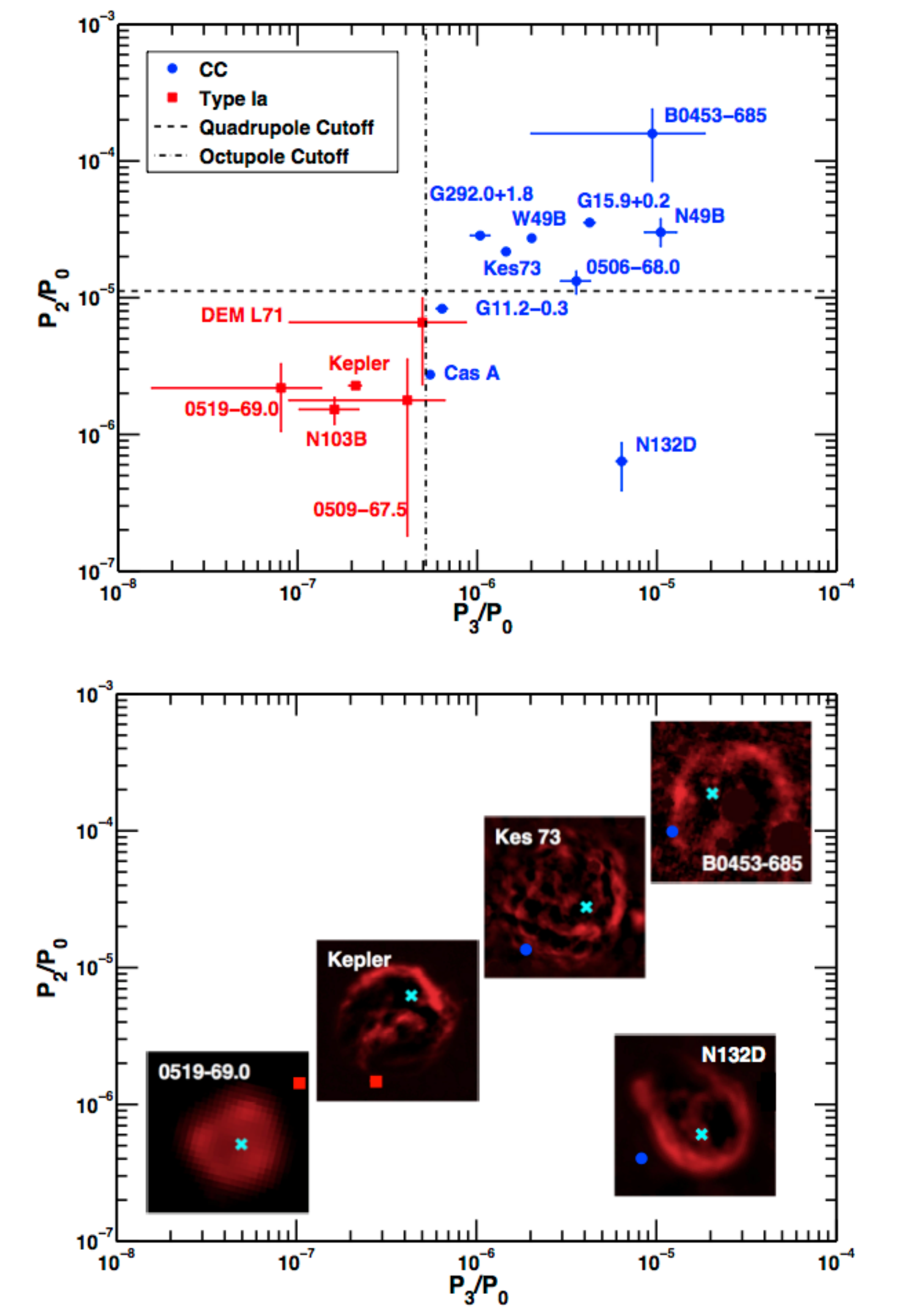}
 \caption{Top: Quadrupole power ratio \ptwo\ vs. octupole power ratio \pthree\ of the 24 $\mu$m emission for the 15 SNRs. The horizontal and vertical lines mark the cutoffs in \ptwo\ and \pthree\, respectively, between the Type Ia SNRs (plotted with red squares) and the CC SNRs (plotted with blue circles). Bottom: 24 $\mu$m images of five SNRs, plotted in the position of their power-ratio values (as identified by the red squares or blue circles). The cyan crosses mark the centroid used to perform the power-ratio method analysis.}
\label{fig:momentsplot}
\end{figure}

The resulting \ptwo\ and \pthree\ values for the MIPS 24 $\mu$m images of our 15 sources are plotted in Figure~\ref{fig:momentsplot}. Overall, we find that the Type Ia and CC SNRs have statistically significant different power ratios. The CC SNRs have a mean $P_{2}/P_{0}$ = ($3.27^{+0.81}_{-0.88}$)$\times 10^{-5}$ and a mean $P_{3}/P_{0}$ = ($3.98^{+0.94}_{-0.77}$)$\times 10^{-6}$; the Type Ia SNRs have a mean $P_{2}/P_{0}$ = ($2.87^{+0.79}_{-0.95}$)$\times 10^{-6}$ and a mean $P_{3}/P_{0}$ = ($2.71^{+0.91}_{-1.05}$)$\times 10^{-7}$. To verify the statistical separation of the Type Ia and CC SNR subsamples, we performed T-tests on their \ptwo\ and \pthree\ values.  The \ptwo\ values of the two subsamples are different at the 99.4\% confidence level, and \pthree\ values are distinct with 99.97\% likelihood. A two-dimensional Kolmogorov-Smirnov (KS) test confirmed these findings, showing the Type Ia and CC SNRs have different power ratios with 99.9\% confidence.

Generally, the CC SNRs populate a distinct part of the power ratio diagram than the Type Ia SNRs, and thus we can define boundaries in \ptwo\ and \pthree\ between the two classes. Most of the CC SNRs have \ptwo $\gs 10^{-5}$ (although a few CC sources are below this limit: N132D, Cas~A, and G11.2$-$0.3), while all of the Type Ia SNRs have \ptwo $\ls 10^{-5}$. All of the CC SNRs have \pthree $\gs 5.0 \times 10^{-7}$, and the Type Ia SNRs all have lower \pthree (although the uncertainty in \pthree\ for DEM~L71 extends into the CC range). We note that the boundaries dividing the two regions are somewhat uncertain given the small number of sources in our sample. However, the typing of individual SNRs of unknown origin based on 24 $\mu$m morphology is likely reliable if the \ptwo\ and \pthree\ values are significantly above or below these dividing boundaries. 

These quantitative results suggest that as a class, Type Ia SNRs have more circular and mirror symmetric IR emission than CC SNRs. These findings complement those from the X-ray analyses of L09a and L11, which showed that the X-ray line and thermal bremsstrahlung emission of Type Ia SNRs is more circular and mirror symmetric than CC SNRs. Although the morphological results in these two wavelength regimes are similar, the IR and X-ray emissions probe distinct emitting regions. In particular, the thermal X-rays trace the ejecta material heated by the reverse shock, while the IR originates from circumstellar dust heated by interaction with the blast wave (see Figures~\ref{fig:24emissionIa} and \ref{fig:24emissionCC} for a comparison of the IR and X-ray images). However, the asymmetries derived from both the X-rays and IR are highly sensitive to the shape of the contact discontinuity, where the reverse-shocked ejecta are impacting the shocked ISM, since emission at larger radii is weighted more heavily in the moment calculation. As the contact discontinuity is shaped by both explosion geometry as well as the structure of the surrounding medium, the distinct symmetries of Type Ia and CC SNRs at 24 $\mu$m reflect the different explosion mechanisms and environments of the two classes.

We note that the sample included in this study is only young-to-middle aged ($\sim$300--13000 year old) SNRs with bright thermal X-ray emission. The IR emission of these SNRs is thought to be dominated by warm dust, but the IR emission in other SNRs may arise from the various mechanisms described in Section \ref{sec:introduction}. For example, most of the IR flux density in plerionic SNRs (those shaped by their pulsar wind nebulae, like the Crab Nebula) comes from synchrotron losses \citep{strom92}, and the morphological properties of these sources may differ from those considered here. Furthermore, older SNRs will eventually ``forget'' their explosive origin as their shocks slow down and become radiative. 

\section{Summary} \label{sec:summary}

In this paper, we have used the power-ratio method (PRM) to quantify the 24 $\mu$m IR morphologies of young SNRs. We have demonstrated that Type Ia SNRs have statistically more circular and mirror symmetric IR emission due to the distinct explosion mechanisms and environments of the two classes of SNe. The ability to distinguish between the two populations offers a new means to constrain SNR explosion types using IR imaging. Specifically, CC SNRs have \ptwo$\gs 10^{-5}$ and \pthree $\gs 5 \times 10^{-7}$. In this wavelength regime, several all-sky surveys have been undertaken, such as the recent Wide-field Infrared Survey Explorer (WISE; \citealt{wright10}), and these data could be employed to explore the explosive origin of other nearby SNRs. Future high-resolution mid-IR imaging, with e.g., the James Webb Space Telescope Mid-Infrared Instrument (MIRI: \citealt{wright04}), may allow the typing of SNRs in more distant Local Group galaxies with this method as well. 

\acknowledgments

The authors would also like to thank Daniel Castro, Rodolfo Montez Jr., Sarah Pearson, and Alicia Soderberg for helpful discussions. This research is based on observations made with the {\it Spitzer} Space Telescope, which is operated by the California Institute of Technology's Jet Propulsion Laboratory under a contract with NASA and made available to the public via the {it\ Spitzer} data archive. Financial support for this research was provided by the National Science Foundation (AST-0849736 and AST-0847563) and the David and Lucille Packard Foundation. Support for LAL was provided by NASA through the Einstein Fellowship Program, grant PF1--120085, and the MIT Pappalardo Fellowship in Physics.

{\it Facilities:} \facility{Spitzer (MIPS)}, \facility{CXO}

\end{document}